\documentclass[english,aps,pre,nofootinbib]{revtex4}
\usepackage[T1]{fontenc}
\usepackage[latin9]{inputenc}
\usepackage{amsmath}
\usepackage{amssymb}
\usepackage{graphicx}
\usepackage{esint}

\makeatletter
\@ifundefined{textcolor}{}
{%
 \definecolor{BLACK}{gray}{0}
 \definecolor{WHITE}{gray}{1}
 \definecolor{RED}{rgb}{1,0,0}
 \definecolor{GREEN}{rgb}{0,1,0}
 \definecolor{BLUE}{rgb}{0,0,1}
 \definecolor{CYAN}{cmyk}{1,0,0,0}
 \definecolor{MAGENTA}{cmyk}{0,1,0,0}
 \definecolor{YELLOW}{cmyk}{0,0,1,0}
 }

\@ifundefined{showcaptionsetup}{}{%
 \PassOptionsToPackage{caption=false}{subfig}}
\usepackage{subfig}
\makeatother

\usepackage{babel}
\begin{document}

\title{Optimal spatiotemporal focusing through complex scattering media}

\author{Jochen Aulbach\footnote[2]{J. Aulbach and A. Bretagne contributed equally to this work.}}

\email{j.aulbach@amolf.nl}

\affiliation{Institut Langevin, ESPCI ParisTech, CNRS, 10 rue Vauquelin, 75231
Paris Cedex 05, France}

\affiliation{FOM Institute for Atomic and Molecular Physics AMOLF, Science Park
113, 1098 XG Amsterdam, The Netherlands }

\author{Alice Bretagne\footnotemark[2]}

\affiliation{Institut Langevin, ESPCI ParisTech, CNRS, 10 rue Vauquelin, 75231
Paris Cedex 05, France}

\author{Mathias Fink}

\affiliation{Institut Langevin, ESPCI ParisTech, CNRS, 10 rue Vauquelin, 75231
Paris Cedex 05, France}

\author{Mickaël Tanter}

\affiliation{Institut Langevin, ESPCI ParisTech, CNRS, 10 rue Vauquelin, 75231
Paris Cedex 05, France}

\author{Arnaud Tourin}

\affiliation{Institut Langevin, ESPCI ParisTech, CNRS, 10 rue Vauquelin, 75231
Paris Cedex 05, France}
\begin{abstract}
We present a new approach for spatiotemporal focusing through
complex scattering media by wave front shaping. Using a nonlinear
feedback signal to shape the incident pulsed wave front, we show
that the limit of a spatiotemporal matched filter can be achieved,
i.e., the wave amplitude at the intended time and focus position is
maximized for a given input energy. It is exactly what is also
achieved with time-reversal. Demonstrated with ultrasound
experiments, our method is generally applicable to all types of
waves.
\end{abstract}
\maketitle

\section{Introduction}

Most imaging systems are based on the ability to focus a wave beam
inside the area of interest. In the context of echographic imaging,
focusing of ultrasound in the human body can be achieved with a
transducer array and electronic delay lines: the same pulsed
waveform is sent from each transducer with the appropriate delay
making all the waveforms converge in synchrony at the desired focus
location. However, that principle is not practicable any more as
soon as the sample thickness becomes larger than the mean free path,
i.e., the average distance between two random scattering events. In
such a strongly scattering medium, it has been shown with ultrasonic
waves that spatiotemporal focusing can be achieved using
time-reversal: a training pulse is first sent from a source located
at the intended focal point, travels through the scattering medium
and is captured at a transducer array, the 'time reversal
mirror'\cite{mathias_timereversed_1997}. The waveforms received on
the time reversal mirror are flipped in time and sent back,
resulting in a wave converging at the desired focus location.
Time-reversal focusing is optimal is the sense that it achieves a
spatiotemporal matched filter, a term well-known from signal
processing; for a given input energy, the amplitude of the pulse at
the focal spot is maximal \cite{tanter2001optimal}. Time-reversal
has also been implemented for microwaves
\cite{geoffroy_focusing_2007}. For optical waves, time-reversal has
not been demonstrated yet, since optical time-reversal mirrors are
challenging to realize.

Therefore in optics a different approach, named 'wave front shaping'
(WFS), has been taken to steer light through strongly scattering
media \cite{m._focusing_2007}. The principle is based on spatial
phase modulation by Spatial Light Modulators (SLM), which enables
one to modulate the complex amplitude of the independent modes of
the incident wave front. The intensity in the selected output mode,
e.g. the focal spot, is used to match the wave front to the
scattering medium by an adaptive algorithm. Owing to their enormous
amount of pixels, state-of-the-art spatial light modulators allow
achieving transmission of substantial amounts of the input energy
\cite{m._universal_2008}.

A big advantage of adaptive wave front shaping approaches lies in
the fact that a direct access to the amplitude at the focus is not
required, such that any type of intensity probe can be used. For
instance the fluorescence from dye molecules allows to focus on
objects hidden inside a complex medium \cite{m._demixing_2008}. The
concept has been applied to other types of waves, such as surface
waves \cite{bergin_activespatial_2011} and ultrasonic waves in the
single scattering regime for medical imaging
\cite{eric_energybased_2009} and MR-guided ultrasonic therapy
\cite{larrat2010mrguided}.

Initially introduced in the monochromatic regime, WFS has recently
been extended to broadband light for spatiotemporal focusing
\cite{jochen_control_2011}. In this work it was shown that spatial
control of the input waves is sufficient to achieve spatiotemporal
control of the scattered waves. The key idea is to use coherence
gating for optimizing a single specific predetermined point in time.
Performing the optimization at a specific time in the speckle
amounts to combine all the scattering paths that have about the same
length. This principle also works efficiently without defining a
focusing time in the case when the temporal delay spread induced by
the scattering medium is comparable to the input pulse duration
\cite{ori_focusing_2011}. In both cases, the missing control of
temporal degrees of freedom is compensated by manipulating spatial
ones. In a complementary way, temporal control of the input light
based on pulse shaping \cite{weiner2000femtosecond} can be used to
achieve spatiotemporal focusing by  spatially localized phase
compensation \cite{j_shaping_2011}. Both approaches exploit the fact
that temporal and spatial degrees of freedom are mutually
convertible in a complex medium \cite{lemoult2009manipulating}.

In the present paper, we demonstrate how to fully control both
spatial and temporal degrees of freedom to achieve optimal
spatiotemporal focusing  for a broadband ultrasonic wave propagating
through a complex multiple scattering medium. Like in the context of
time-reversal focusing, we use the term 'optimal' in the sense of a
spatiotemporal matched filter, i.e., for a given input energy, our
method maximizes the wave amplitude at the intended time and focus
position. The difference with time reversal is that no amplitude
measurement is required at the focus location. Instead spatial and
frequency resolved wave front shaping can be achieved with a
nonlinear feedback intensity signal and is capable to reach optimal
spatiotemporal focusing.Thus our concept should be generalizable to
all types of waves.

\section{The spatiotemporal matched filter approach}

\subsection{Optimal focusing by time reversal}

\begin{figure}
\subfloat{\includegraphics[width=0.4\textwidth]{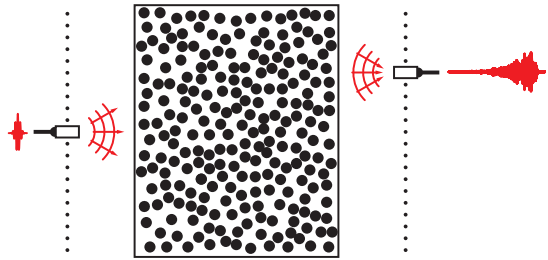}}\caption{\label{fig:experiment}Experimental system. From $N$ points in the
emitter plane on the left, broadband signals are transmitted through
a two-dimensional multiple scattering medium and recorded at $M$
points in the receiver plane on the right. Details of the experiment
are specified in section \ref{sec:Experiment}.}
\end{figure}

The experimental system is shown in Fig.~\ref{fig:experiment}. Our
goal is to maximize the signal amplitude at some focus point $m_{0}$
in the receiver plane at time $t_{0}$ for a given energy radiated
from $N$ points in the emitter plane. In signal processing, this
is known as the spatiotemporal matched filter approach.

\paragraph{Spatial matched filter}

At first we recapitulate the conditions for optimal spatial focusing
in the sense of a spatial matched filter\cite{tanter2001optimal}.
Let $(E(\omega))_{n},1\leqq n\leqq N$ be the input signals in the
emitter plane, and correspondingly $(F(\omega))_{m}$ the signals
received in the image plane. The propagation through the medium from
the emitter points to the control points at one frequency is
described by:
\begin{equation}
F(\omega)=\mathbf{T}(\omega)E(\omega),\label{eq:Tmatrix}
\end{equation}
where $\mathbf{T}(\omega)$ is the propagation operator, or transfer
matrix, between the points at a given frequency $\omega$. The amplitude
received at the focal point $m_{0}$ at a single frequency $\omega$
is given by
\begin{equation}
F_{m_{0}}(E)=\frac{\left\langle \mathbf{T}E|F^{(0)}\right\rangle }{||E||}=\frac{\left\langle E|^{t}\mathbf{T}^{*}F^{(0)}\right\rangle }{||E||},
\end{equation}
where
\begin{equation}
F^{(0)}=\left\{ 0,...,0,1,0,....,0\right\}
\end{equation}
is the projection onto $m_{0}$. The inequality of Cauchy-Schwartz
sets the upper bound for this expression:
\begin{equation}
\frac{\left\langle E|^{t}\mathbf{T}^{*}F^{(0)}\right\rangle }{||E||}\leqq\left\Vert ^{t}\mathbf{T}^{*}F^{(0)}\right\Vert \label{eq:Inequality}
\end{equation}
Equality holds, if $E\propto^{t}\mathbf{T}^{*}F^{(0)}$, or
\begin{equation}
\forall n\; E_{n}(\omega)=\alpha(\omega)T_{m_{0}n}^{*}(\omega).\label{eq:MatchedFilter}
\end{equation}
The maximum amplitude at the focal point $m_{0}$ is reached when
all $E_{n}$ are proportional to the complex conjugate of the respective
transmission coefficient. This situation corresponds to a phase conjugation
experiment or a time-reversal experiment at a single frequency.

\paragraph{Temporal matched filter}

Let us denote by $T_{m_{0}n}(t)$ the impulse response between emitter
$n$ and receiver $m_{0}$, which is the Fourier Transform of the
respective transfer function in the frequency domain $T_{m_{0}n}(\omega)$.
When one aims at maximizing the amplitude received at time $t_{0}=0$,
it is well-known that $E_{n}(t)=T_{m_{0}n}(-t)$ has to be sent from
$n$. The signal at $m_{0}$ then writes as
\begin{equation}
F_{m_{0}}(t)=\sum_{n}T_{m_{0}n}(t)\otimes T_{m_{0}n}(-t)=\sum_{n}\int d\omega\,e^{i\omega t}T_{m_{0}n}(\omega)T_{m_{0}n}^{*}(\omega)\label{eq:Focus}
\end{equation}
which is the autocorrelation of the impulse response, or the inverse
Fourier transform of the power spectrum, respectively. It is an even
function with its maximum value at $t=0$ when all frequency
contributions add in phase. Hence the magnitude at $t_{0}$ is directly proportional to the transmitted energy. As $T_{m_{0}n}(-t)$ is the
Fourier Transform of $T_{m_{0}n}^{*}(\omega)$, it also achieves the
spatial matched filter (Eq.~\ref{eq:MatchedFilter}).

This signal exactly corresponds to the one transmitted in a broadband
time-reversal experiment. In the first step, the control point $m_{0}$
in the image plane behaves like a source, such that the wave field
recorded by the two-way transducers in the emitter plane is given
by $T_{nm_{0}}(t).$ In the second step, the recorded fields are emitted
in a time-reversed manner. Under assumption of reciprocity the emission
writes as $E_{n}(t)=T_{nm_{0}}(-t)=T_{m_{0}n}(-t)$ and thus the focus is given by Eq. \ref{eq:Focus}. Hence, time-reversal exactly achieves
a spatiotemporal matched filter since it implements a spatial filter at
each frequency of the incident pulse and the correct phase relation between
the frequency components is intrinsically recovered.

\subsection{Optimal focusing by wave front shaping}

In the following we lay out the wave front shaping algorithm and investigate
its performance as a matched filter. In order to achieve optimal spatiotemporal
focusing of a broadband pulse, we perform the wave front shaping algorithm
in two stages. During the first phase, we achieve optimal spatial focusing.
Based on this result, optimal temporal focusing is achieved after the second phase.

\subsubsection{Optimal spatial focusing\label{sub:Optimal-spatial-focusing}}

The direct implementation of wave front shaping as applied in optics
to monochromatic acoustic waves would be described as follows: at
first an arbitrary wave front is sent from the emitter plane. The
intensity is then measured at the intended focal point. This can be
done directly in measuring the pressure field with a hydrophone or,
in the case of a soft elastic medium, in measuring the displacement
induced at the focal point by the radiation force, which is proportional
to the acoustic intensity \cite{eric_energybased_2009}. Then the
phase of the first transducer is cycled between $0$ and $2\pi$ while
the intensity is recorded. The phase value for which intensity at
focus is found maximum is then stored in the memory and the same operation
is repeated for each transducer. Once the optimal phase has been stored
for each transducer, the wave front can be synthesized by exciting
each transducer with the determined optimal phase.

\paragraph{Hadamard method\label{par:Hadamard}}

Instead of using such an optimization scheme element by element, we
perform a basis transformation to construct virtual transmitters $\hat{E}_{j}$,
defined as linear combinations of real ones, which greatly improves
the sensitivity of the method and reduces the number or iterations
\cite{eric_energybased_2009}. The coefficients of the combination
are taken as the columns $H_{n}(1<n<N)$ of an $N$ by $N$ Hadamard
matrix $H$, with elements $H_{jn}\epsilon\left\{ -1;1\right\} $.
This choice ensures that the amplitude transmitted from each transducer
is maximal. The basis transformation is performed by
\begin{equation}
\hat{E}=\mathbf{H^{-1}}E
\end{equation}
and the inverse transformation by
\begin{equation}
E=\mathbf{H}\hat{E}.
\end{equation}
The field in the receiver plane is given by
\begin{equation}
F=\mathbf{T}E=\mathbf{T}\mathbf{H}\hat{E}=\mathbf{\hat{T}}\hat{E}
\end{equation}
where $\mathbf{\hat{T}}$ is the transfer matrix between the emitters
in the hadamard basis and the receivers in the canonical basis. For
the optimization algorithm, we arbitraritly chose the first column
as the reference. 4$N$ coded beams are sent into the scattering medium
as:

\begin{equation}
C_{j,\delta}=\hat{E}_{ref}+\hat{E}_{j}e^{i\delta}
\end{equation}

\begin{equation}
j\epsilon[1,N];\;\delta\epsilon[0,\pi,\frac{\pi}{2},-\frac{\pi}{2}]
\end{equation}
After each transmission the resulting intensity at the focal point
$m_{0}$ is given by
\begin{equation}
I_{j,\delta}=a_{ref}^{2}+a_{j}^{2}+2a_{ref}a_{j}\cos(\phi_{ref}-\phi_{j}-\delta),\label{eq:HadamardInt}
\end{equation}
where $a_{j}e^{i\phi_{j}}=\hat{T}_{m_{0}j}$ is the corresponding
transmission coefficient from the transfer matrix $\hat{\mathbf{T}}$.
From the set of measured intensities we are now able to calculate
the optimal wave front. The complex amplitude to be transmitted from
each virtual transducer to optimize the intensity is given by the
following expressions:
\begin{equation}
\hat{E}_{j}^{smf}=a_{j}^{smf}e^{i\phi_{j}^{smf}}\left\{ \begin{array}{c}
a_{j}^{smf}=a_{j}=\sqrt{I_{j,0}+I_{j,\pi}-\frac{I_{ref,0}}{2}}\\
\phi_{j}^{smf}=\phi_{ref}-\phi_{j}=\arctan(\frac{I_{j,\pi/2}-I_{j,-\pi/2}}{I_{j,0}-j_{n,\pi}})
\end{array}\right.\label{eq:HadamardCalc}
\end{equation}
where the $\phi_{j}^{smf}$ compensate for the phase differences between
the contributions at the focal point. The choice of $a_{j}^{smf}$
ensures that WFS achieves a spatial matched filter. Consequently,
all $\hat{E}_{j}^{smf}$are equivalent to their corresponding $\hat{T}_{m_{0}j}^{*}$
up to the global phase difference $\phi_{ref}$,
\begin{equation}
\hat{E}^{smf}=e^{i\phi_{ref}}(\hat{\mathbf{T}}^{*})^{t}F^{(0)},\label{eq:WfsOptimalEmissionFullHada}
\end{equation}
where $F^{(0)}$ selects the $m_{0}$-th row of the transmission
matrix. Finally, a change of basis leads back to the transducer
basis in the emitter plane
\begin{equation}
E^{smf}=\mathbf{H}\hat{E}^{smf}=e^{i\phi_{ref}}\mathbf{H}(\hat{\mathbf{T}}^{*})^{t}F^{(0)}=Ne^{i\phi_{ref}}(\mathbf{T}^{*})^{t}F^{(0)},\label{eq:WfsOptimalEmission}
\end{equation}
showing, that the wave front shaping algorithm fulfils the condition
of the spatially matched filter (Eq.~\ref{eq:MatchedFilter}). The
emission is equivalent to the time reversal experiment.
With the definition of the Hadamard matrix, $\mathbf{HH^{t}=\mathrm{N}1}$,
the prefactor N arises since we omitted the equivalent transformation
in the receiver plane. Since this proportionality factor does not
impair any conclusions (see Eq.~\ref{eq:MatchedFilter}), we omit it
in the following.

\paragraph{Enhancement factor}

After optimization for focusing on point $\rm{m_0}$, the emitted signal from
transducer $n$ is
\begin{equation}
E_{n}^{smf}=\frac{T_{m_{0}n}^{*}}{\sqrt{\sum_{n}\left|T_{m_{0}n}\right|^{2}}},\label{eq:WfsEmission}
\end{equation}
where the denominator normalizes the transmission. The received signal
at focus point $m_{0}$ is
\begin{equation}
F_{m_{0}}=\frac{\sum_{n}T_{m_{0}n}T_{m_{0}n}^{*}}{\sqrt{\sum_{n}\left|T_{m_{0}n}\right|^{2}}}=\sqrt{\sum_{n}\left|T_{m_{0}n}\right|^{2}}.
\end{equation}
The average energy received at the focus point $m_{0}$ is
\begin{equation}
J{}^{(1)}=\left\langle \left|F_{m_{0}}\right|^{2}\right\rangle
=\left\langle \sum_{n}\left|T_{m_{0}n}\right|^{2}\right\rangle
=N\left\langle |T_{m_0n}|^{2}\right\rangle,
\end{equation}
where the brackets denote the statistical average. We assume that
the transmission coefficients are independent random variables and
follow a circular Gaussian distribution. Without optimization, the
normalized initial emission is
\begin{equation}
E_{n}=\frac{1}{\sqrt{\sum_{n}1^{2}}}=\frac{1}{\sqrt{N}}.
\end{equation}
The average received signal on point $m_{0}$ is
\begin{equation}
J{}^{(0)}=\left\langle \left|F_{m_{0}}\right|^{2}\right\rangle
=\left\langle
\frac{\sum_{n}T_{m_{0}n}\sum_{n'}T_{m_{0}n'}^{*}}{\left(\sqrt{N}\right)^{2}}\right\rangle
=\left\langle \left|T_{m_0n}\right|^{2}\right\rangle.
\end{equation}
The average enhancement, which also gives the signal to background
ratio to other non-optimized modes, is therefore given by
\begin{equation}
\eta=\frac{J{}^{(1)}}{J{}^{(0)}}=N.\label{eq:Enhancement}
\end{equation}
Initial works in optics \cite{m._focusing_2007} used spatial light
modulators which were limited to phase-only control of the emitted
amplitudes, such that the emission is given by
\begin{equation}
E_{n}^{po}=\frac{1}{\sqrt{N}}\frac{T_{m_{0}n}^{*}}{\left|T_{m_{0}n}\right|}.\label{eq:WfsEmissionPhaseOnly}
\end{equation}
The amplitude at the focus is
\begin{equation}
F_{m_{0}}=\frac{1}{\sqrt{N}}\sum_{n}\left|T_{m_{0}n}\right|.
\end{equation}
In this case the received energy at focus point m is
\begin{equation}
J{}^{(2)} =\left\langle \left|F_{m_{0}}\right|^{2}\right\rangle = \left\langle \left|T_{m_0n}\right|^{2}\right\rangle
+(N-1)\left\langle \left|T_{m_0n}\right|\right\rangle ^{2}
\end{equation}
Hence the enhancement is lowered by the well-known prefactor
\begin{equation}
\frac{J{}^{(2)}}{J{}^{(1)}}\approx\frac{\left\langle
\left|T_{m_0n}\right|\right\rangle ^{2}}{\left\langle
\left|T_{m_0n}\right|^{2}\right\rangle
}=\frac{\pi}{4}\approx0.8.\label{eq:EnhancementPrefactor}
\end{equation}

\paragraph{Algorithm for optimal spatial focusing}

During the first phase, we achieve spatial focusing for all frequency
components contained in emission spectrum of the broadband pulse with its bandwidth $\Delta\omega_{bb}$. Since the transmission
coefficients $T_{nm}(\omega)$ (Eq.~\ref{eq:Tmatrix}) are random
variables of frequency with a correlation length $\delta\omega<\Delta\omega_{bb}$,
the optimal emission $E^{smf}(\omega)$ (Eq.~\ref{eq:WfsOptimalEmission})
has to be determined frequency-resolved for intervals $\Delta\omega<\delta\omega$.
Consequently, the Hadamard algorithm is performed frequency by frequency
in steps of $\Delta\omega$, each time using a narrowband pulse with
a Gaussian spectrum of a bandwidth $\Delta\omega$ for the emission.
When afterwards the full broadband signal with the optimal coefficients
$E^{smf}(\omega)$ is emitted, spatially matched focusing is achieved.
Contrary to the time-reversal experiment, temporal focusing can not
yet be expected. For this, all frequency components have to be in
phase, which is intrinsically fulfilled by time reversal. However,
after the wave front shaping optimization, the unknown phase factor $e^{i\phi_{ref}(\omega)}$
remains.

\subsubsection{Optimal temporal focusing\label{sub:Optimal-temporal-focusing}}

During the second phase, the spatially focused signal is additionally
focused in the time-domain. For that, the phase factor $e^{i\phi_{ref}(\omega)}$
needs to be determined and its conjugate equally multiplied to all
emissions $E_{n}^{smf}(\omega)$. Such spatially invariant phase filtering
leads only to a temporal redistribution of the signal at the focal
point. The time-integrated intensity at the focus is unaltered by
the filtering process, excluding it as an appropriate feedback signal.
Instead of detecting the linear intensity at the focal spot, we place
a nonlinear detector which allows us to measure the time-integrated
second harmonic intensity. This detector is treated in more details
in the following paragraph. The second harmonic energy is sensitive
for a temporal redistribution of the pulse energy. Used as a feedback
signal, it reaches its maximum when the transmitted signal is shortest
in duration, corresponding to the so-called Fourier-limit, when the
waves at all frequencies are in phases at the focus. This technique
has similarly been used in optics for the compression of dispersed
ultrashort laser pulses \cite{d._adaptive_1997}.

The most straight-forward approach is an adaptive algorithm
equivalent to the spatial optimization scheme described in the
beginning of the section. The frequency range of the broadband emission spectrum is subdivided into intervals $\Delta\omega<\delta\omega$.
Within each iteration $j$ of the optimization, we emit the full signal
as determined for optimal spatial focusing, $E_{n}^{smf}$, but modified with a
spatial invariant phase mask $e^{i\Delta\phi(\omega)}$. Step by
step, all frequency intervals are addressed successively. Before the
first iteration, the phase mask is
$\Delta\phi(\omega)=0\,\forall\omega$. In step $j$, the phase mask
is modified only at the frequency $\omega_{j}$. We perform several
emissions with the phase $\Delta\phi(\omega_{j})$ increasing from
$\Delta\phi(\omega_{j})=0$ to $\Delta\phi(\omega_{j})=2\pi$, while
the second harmonic energy $J_{j}^{SH}$ is recorded. As a function
of each $\Delta\phi(\omega_{j})$, $J_{j}^{SH}$ follows a cosine
behaviour
$J_{j}^{SH}\propto\cos[\Delta\phi(\omega_{j})+\phi_{SH}(\omega)]$
with a phase offset $\phi_{SH}(\omega_{j})$, which depends on the
actual phases of all other frequency components. The optimum phase
offset $\Delta\phi(\omega_{j})=-\phi_{SH}(\omega_{j})$ is found when
$J_j^{SH}$ reaches its maximum. The modified phase mask is subsequently
used in the next step address, in which the next frequency interval
$\omega_{j+1}=\omega_{j}+\Delta\omega$ is handled in the same way.
This way, the independent frequency components of the signal are set
in phase at the focus step by step, i.e. the phase mask
$e^{i\Delta\phi(\omega)}$ converges towards the conjugate of
$e^{i\phi_{ref}(\omega)}$.

\subsubsection{Nonlinear detector}

For the first phase, to achieve spatial focusing, the detection of
the linear intensity would be sufficient. However, for an experimentally
convenient approach, we use the nonlinear detector for both spatial
and temporal focusing. We consider a system with a second-order nonlinearity,
which could be realized, e.g., by a nonlinear scatterer, such as microbubbles in acoustics \cite{g._theacoustic_1994}
or nonlinear nanocrystals \cite{chia-lung_threedimensional_2009}
in optics, combined with a time-integrated detection of the scattered second-order
response. We calculate the detector response by
\begin{equation}
J^{SH}=\alpha\int dt|F_{m_{0}}(t){}^{2}|^{2},
\end{equation}
where $\alpha$ is an arbitrary prefactor and $F_{m_{0}}(t)$ the
amplitude at the focus point in the time domain. For the narrowband
pulses emitted in the first phase, we can assume that the linear intensity
can be readily obtained by $I=\sqrt{J^{SH}}$ for the calculation
of the transmission coefficients (Eq.~\ref{eq:HadamardCalc}).

\section{Experiment\label{sec:Experiment}}

\subsection{Transfer matrix measurement}

As a first step of the experiment, we recorded the transfer matrices
of a multiple scattering medium over a broad range of frequencies.
The experimental setup consists of two identical transducers, one
used as emitter and the other as receiver respectively, placed in
water on opposite sides of the medium (Fig.~\ref{fig:experiment}). The latter is an arrangement (thickness L = 80 mm) of parallel steel
rods with density $18\,\mathrm{cm}^{-2}$ and a rod diameter 0.8 mm.
Both the emitting and the receiving transducer can be translated on
a line parallel to the medium, on which we defined each 128 emitter
points and 128 receiver points spaced by 0.4 mm. From each emitter
point, we sent an ultrasonic pulse at a central frequency of 3.5 MHz
with a relative bandwidth of 31\%. In this frequency range the mean
free path of the medium is $l\approx4\mathrm{mm}$ as determined in
\cite{arnaud_randommultiple_2001}. The transmitted signals were recorded
on all receiver points with a sampling frequency of 10 MHz. From the
Fourier transform of the temporal signal, we obtain the transfer matrices
in the considered frequency range (Eq.~\ref{eq:Tmatrix}). We determined
the correlation of the transmission coefficients both in the spatial
and in the frequency domain by
\begin{equation}
C_{\omega}(\Delta\omega)=\left\langle
T_{nm}(\omega)T_{nm}^{*}(\omega+\triangle\omega)\right\rangle
_{n,m},\label{eq:Cfreq}
\end{equation}
\begin{equation}
C_{r}(\Delta r)=\left\langle T_{nm}(\omega)T_{n(m+\Delta
r)}^{*}(\omega)\right\rangle _{n,\omega},\label{eq:Cspatial}
\end{equation}
where the average is performed over the denoted matrix coefficients in Eq.~\ref{eq:Cfreq} and over the first matrix entry and frequencies in Eq.~\ref{eq:Cspatial}.
From the full width at half
maximum of the correlation functions,
$\mathrm{\delta\omega=FWHM}(\left|C_{\omega}\right|)=5.14\mathrm{kHz}$
and $\mathrm{FWHM}(\left|C_{r}\right|)=1.04\mathrm{mm}$, we can
determine that we have $1/5.14\mathrm{kHz}$ independent frequencies
per frequency unit and
$N=(128\cdot0.40\mathrm{mm})/1.04\mathrm{mm}=49.2$ independent
emitter points available \cite{frerik_frequency_2011}.

With the transfer matrices at hand, we can calculate the linear
response of the system for arbitrary signals emitted either from the
'emitter' or 'transducer' side. For all following experiments, we
emitted pulses with a Gaussian spectral function of 10\% bandwidth
around the central frequency from all transducers. A typical signal
in the receiver plane is plotted in Fig.~\ref{fig:Received-Maps}(a) and Fig.~\ref{fig:ReceivedOnFocus}(a). The energy is
spread widely both temporally and spatially. The ballistic part of
the wave has disappeared, which confirms that the medium is strongly
scattering.

\subsection{Spatial focusing}

\begin{figure}
\subfloat{\includegraphics[width=0.7\textwidth]{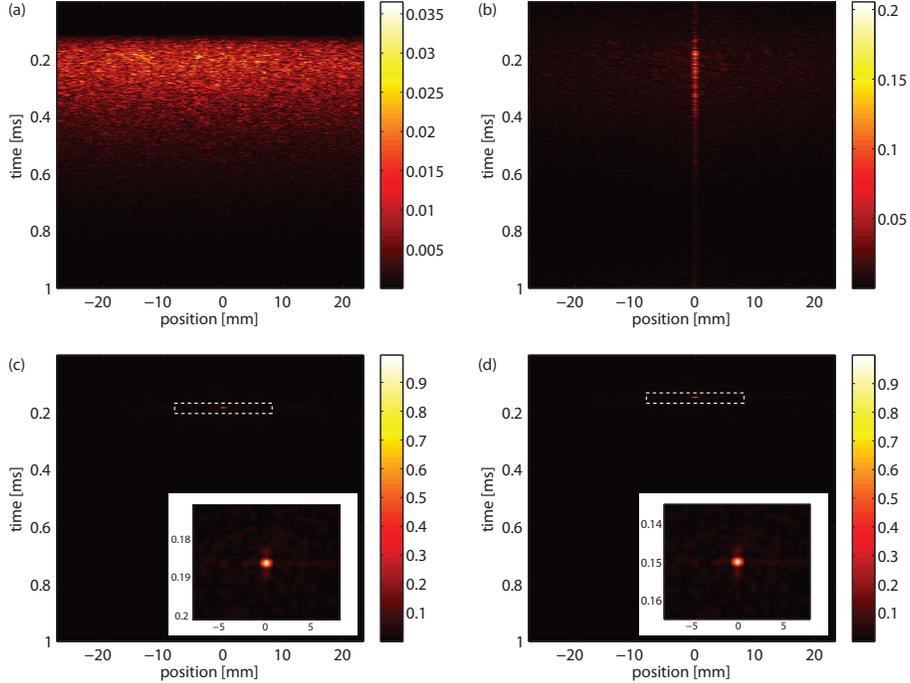}}

\caption{Received amplitude for nonoptimized emission (a), for spatial
wave front shaping (b) and subsequent temporal focusing (c), and for comparison, for the time-reversal focus (d).
The areas in the dashed rectangles are plotted magnified in the inlays.\label{fig:Received-Maps}}
\end{figure}

\begin{figure}
\subfloat{\includegraphics[width=0.4\textwidth]{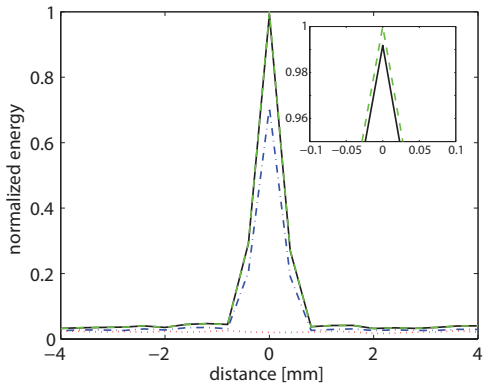}}

\caption{Received energy before (dotted line), after wave front shaping with
phase-only control (dash-dotted line) and combined control of amplitude
and phase (solid line). The focal spot obtained with time-reversal
is plotted for comparison (dashed line). \label{fig:Energy}}

\end{figure}

Starting from this emission we performed a frequency-resolved
wavefront shaping optimization as described in section
(\ref{sub:Optimal-spatial-focusing}). The bandwidth of the
narrowband pulses is chosen to $\Delta\omega=1.4\,\mathrm{kHz}$ to be
smaller than the correlation $\delta\omega$. In two experiments we
use both emission with optimal phase and amplitude
(Eq.~\ref{eq:WfsEmission}) and the emission with optimal phase only
(Eq.~\ref{eq:WfsEmissionPhaseOnly}). The deposed energy in the
receiver plane for both experiments is plotted in Fig.~\ref{fig:Energy}. It is calculated by integrating over time the
square of the wave amplitude at the receiver points (see Fig.~\ref{fig:Received-Maps}(b) for the signal obtained in optimizing both
amplitude and phase). The optimization leads to a spatial focusing
in both cases. For phase and amplitude control, the increase in
energy at the focal spot is $\eta=49.2$, which corresponds perfectly
to the number of estimated independent emitters $N=49.2$
(Eq.~\ref{eq:Enhancement}). The enhancement by the time-reversal
focus is slightly higher, indicating remaining deviations between
the wave front shaping emission and the time-reversal emission. In
further calculations, we observed that a further reduction of
$\Delta\omega$ eliminates these deviations, which shows that the
single-frequency TR emission and the WFS emission are identical (see Eq.~\ref{eq:WfsEmission} and Eq.~\ref{eq:MatchedFilter}).

\begin{figure}
\subfloat{\includegraphics[width=0.7\textwidth]{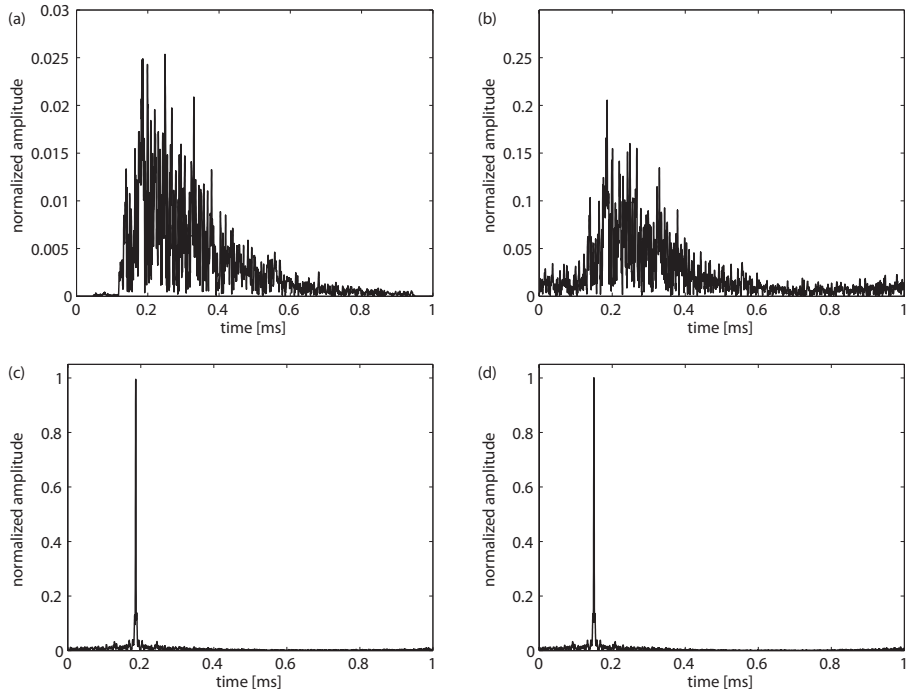}}

\caption{Received amplitude at the focal point for nonoptimized emission (a), for spatial wave front shaping (b) and subsequent temporal
focusing (c), and for comparison, for time-reversal focusing
(d). \label{fig:ReceivedOnFocus}}
\end{figure}

For phase-only optimization, we observe an enhancement lowered by a factor 0.71
($\eta_{po}=35.1$). The factor $0.8$
(Eq.~\ref{eq:EnhancementPrefactor}) is not fully reached, which we
attribute to the fact, that due to the geometry of the setup not all
emitters contribute equally at the focal point, effectively reducing
the number of emitter points \cite{jochen_control_2011}. The
resulting time-resolved signals for the first case are shown in Fig.~\ref{fig:Received-Maps}(b) and Fig.~\ref{fig:ReceivedOnFocus}(b). The energy is still spread
temporally, since temporal focusing is impeded by the remaining
random phase relation between independent frequency components
(Eq.~\ref{eq:WfsOptimalEmissionFullHada}). In the next step, we
determine and correct for this phase factor.


\subsection{Spatiotemporal focusing}

Starting with the emission for spatial optimization, we perform
frequency phase filtering as described in section
(\ref{sub:Optimal-temporal-focusing}). After two cycles over all
frequencies, the received signals are not only spatially, but also
temporally focused (Fig.~\ref{fig:Received-Maps}(c) and
Fig.~\ref{fig:ReceivedOnFocus}(c)). The resulting peak
intensity can be put in context with the degrees of freedom
available for the optimization. We have $N_{spatial}=49.2$
independent emitters at hand. In the frequency domain,
$N_{freq}=96.3$ degrees of freedom are available calculated from the
frequency correlation, $\delta\omega=5.14\mathrm{kHz}$ and a
bandwidth of $\triangle\omega_{bb}=495\mathrm{kHz}$. Hence, the
total number of degrees of freedom is
$\eta_{theory}=N_{spatial}\cdot N_{freq}=4.7\cdot10^{3}$. We
estimate the intensity enhancement to $\eta_{exp}=6.2\cdot10^{3}$,
comparing the peak intensity to the intensity before optimization
(obtained from the transmission averaged over the transducers around
the focus around the peak of the diffuse transmission, smoothed by a
$60\mathrm{\mu s}$ window to remove temporal speckle). This number
corresponds well to the order of magnitude of the number of degrees
of freedom.

\begin{figure}
\subfloat{\includegraphics[width=0.4\textwidth]{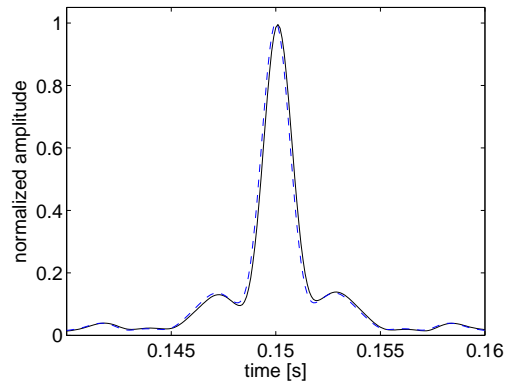}

}

\caption{Amplitude after wave front shaping (solid line) and time reversal
(dashed line) on the focal spot. \label{fig:Comparison}}

\end{figure}

As a reference, we performed an equivalent time-reversal experiment,
which intrinsically reaches spatiotemporal focusing (Fig.~\ref{fig:Received-Maps}(d) and Fig.~\ref{fig:ReceivedOnFocus}(d)). The quality of the
focusing from our adaptive wave front shaping and time-reversal are
nearly identical (Fig.~\ref{fig:Comparison}). The small deviations
are a result of small remaining phase differences which can be
minimized by a reduction of the frequency steps during the spatial
optimization (see above) and further iterations of the adaptive
phase filtering process. As an alternative to the step-by-step
algorithm we used here, so-called stimulated annealing algorithms
\cite{d._adaptive_1997} or genetic algorithms
\cite{t._feedbackcontrolled_1999} should equivalently find the
optimal phase relation. These algorithms are known to be robust in
the presence of noise, but have the disadvantage to turn inefficient
for an increasing number of degrees of freedom.

\section{Conclusions}

In conclusion, we have presented a new approach for spatiotemporal
focusing through complex scattering media by feedback-based
wavefront shaping. In contrast to previous works, our approach is
capable of reaching the limit of a spatiotemporal matched filter;
the maximum possible amount of the emitted energy is deposed at the
target. We showed that to achieve this, phase and amplitude of the
emission need to be controlled at each frequency. In contrast to the
classical time reversal experiment that achieves the ideal focus,
here the direct access to the field amplitude at the target is not
required as only the intensity associated to a nonlinear response
needs to be detected. This point of is particular interest for many
applications in wave physics where the accessible information is
restricted to field intensity. For example, in optics, our method
can be realized by fluorescent dye molecules as they are used for
two-photon microscopy. In ultrasound therapy, it could optimize both
spatial and temporal focusing of therapeutic beam through the skull
bone thanks to MR radiation force imaging. Being generally
applicable to all types of waves, we believe that our method is
promising for a wide range of applications in imaging and sensing,
and for the control of wave propagation in combination with complex
media such as new metamaterials.

This work is part of the Industrial Partnership Programme (IPP) Innovatie
Physics for Oil and Gas (iPOG) of the Stichting voor Fundamenteel
Onderzoek der Materie (FOM), which is supported financially by Nederlandse
Organisatie voor Wetenschappelijk Onderzoek (NWO). The IPP MFCL is
co-financed by Stichting Shell Research.


\end{document}